\documentclass[preprint,review,12pt]{elsarticle}

\usepackage{amssymb} 
\usepackage{amsmath}
\usepackage{graphicx}
\usepackage{subfigure}
\usepackage{bm} 
\usepackage{verbatim} 
\usepackage{color} 

\journal{Physica D: Nonlinear Phenomena}

\begin{document}
\begin{frontmatter}

\title{Power Network Uniqueness and Synchronization Stability from a Higher-order Structure Perspective}

\author[chen]{Hao Liu}
\author[chen]{Xin Chen\corref{cor1}}
\author[chen]{Long Huo}
\author[chen]{Chunming Niu}
\cortext[cor1]{E-mail: xin.chen.nj@xjtu.edu.cn}

\address{1. Center of Nanomaterials for Renewable Energy, State Key Laboratory of Electrical Insulation and Power Equipment, School of Electrical Engineering, Xi'an Jiaotong University}

\begin{abstract}
Triadic subgraph analysis reveals the structural features in power networks based on higher-order connectivity patterns.
Power networks have a unique triad significance profile (TSP) of the five unidirectional triadic subgraphs in comparison with the scale-free, small-world and random networks. 
Notably, the triadic closure has the highest significance in power networks. Thus, the unique TSP can serve as a structural identifier to differentiate power networks from other complex networks. Power networks form a network superfamily.
Furthermore, synthetic power networks based on the random growth model grow up to be networks belonging to the superfamily with a fewer number of transmission lines.
The significance of triadic closures strongly correlates with the construction cost measured by network redundancy. The trade-off between the synchronization stability and the construction cost leads to the power network superfamily. The power network characterized by the unique TSP is  the consequence of the trade-off essentially. The uniqueness of the power network superfamily tells an important fact that power networks maintain a high level of synchronization stability at a low construction cost.
\end{abstract}

\begin{highlights}
	\item We find a unique structural feature based on the triad significance profile in power networks 
	\item The triad significance profile of power grids is unique compared to other networks
	\item Synthetic networks grow up to be power networks with a fewer number of transmission lines
	\item Power networks maintain a high level of synchronization stability at a low construction cost
\end{highlights}


\end{frontmatter}

\section{Introduction}
A power system generally consists of power stations, electrical substations, transmission lines, and electricity consumers. All of these components are linked together. Therefore, a power system can be naturally described as a complex network.  
The complex network analysis provides valuable insights into the characteristics of power system structures and dynamics. For instance, it is used to rank the importance of components\cite{Lu2016}, detect clusters\cite{Newman2006} and recognize global structural features of networks\cite{Watts1998,Barabasi1999,Estrada2010}. But most previous studies on characteristics of power system structures and dynamics are investigated by lower-order properties based on single nodes and edges. Power grids exhibit rich, higher-order connectivity in the real world. Establishing the relationship between higher-order structures and dynamics for power grids has great theoretical value and practical significance. 

The higher-order structures of networks arise from the local functional unit in complex systems. 
Network motifs, the small repeated subgraphs, are considered as the building blocks of complex networks. 
The higher-order structural features of networks can be found\cite{Benson2016a,Tsourakakis2016} more effectively at the level of motifs than at the level of individual nodes and edges. Moreover, the subgraphs are organized in a particular way to form complex networks so that the organization of subgraphs is used to characterize different types of networks\cite{Milo2004a}. 
Understanding the functional role of the higher-order structures in power networks is an open problem, \textit{i.e.}, recent studies show that the motif-based network models can bring new insights into the complicated topology-dynamics relation\cite{Lambiotte2019,Lucas2020} beyond the node-based or edge-based network models. 

Previous studies mainly focus on two aspects of the topology-dynamics relation in power systems: stability of frequency synchronization and robustness against cascading failures.  
Dynamic models estimate the synchronization stability. The power grids are considered to be coupled oscillator networks whose dynamics are governed by the swing equation. Some distinct higher-order structures are identified to emphasize the important role of the local structures in synchronization. For example, the tree-like local structures\cite{Menck2014a} can strongly diminish synchronization stability. On the other hand, the nodes in triangles with low betweenness, called detour nodes\cite{Schultz2014a}, often own high stability. Furthermore, the modular structure of a power grid affects the distribution of the operational resilience of nodes\cite{Kim2021}. 
However, it should be mentioned that the special performance of higher-order structures is highly influenced in the context of global network structures, \textit{i.e.}, the global synchronization of complex networks may be reduced when networks become more clustered with an increasing number of triangles\cite{McGraw2005}. Also, the basin stability transitions with the increasing coupling strength for all 4-node and 6-node motif structures are studied without the context of networks\cite{Kim2016}. But in the context of networks, it is difficult to discuss the functional role of the subgraphs\cite{Gorochowski2018}. 

Quasi-static models generally estimate robustness against cascading failures. The power grids are considered pure complex networks whose robustness is calculated based on complex network theory and DC/AC models\cite{Cuadra2015}. The local subgraph structures are identified to study the network robustness. For example, network motifs can be used as a warning signal for the higher risk of large outages under continuous line/node removal scenario\cite{Chen2018}, and motif concentrations are potentially used as alternative local metrics of robustness under attacks\cite{Dey2019,Dey2018}.
The real-world power grids evolve according to the engineering guidance and standards\cite{Seifi2011} so that there are unique structural features in their networks.
However, to our knowledge, few studies are focused on the relationship between the distinct motif-based topological features of power grids in the real world and their impacts on dynamics. Although the robustness against cascading failures has been well studied based on the quasi-static model from a subgraph perspective\cite{Chen2018,Dey2019,Dey2018}, we aim to reveal the hidden mechanism of synchronization stability and the high-order structures based on the swing equation.

In this paper, we use the significance profile of triadic subgraphs to identify the higher-order structural features and find that power networks have a unique triad significance profile (TSP) compared to the scale-free, small-world and random networks.   
We use the random growth model\cite{Schultz2014}, which is especially proposed to describe the expanding of power grids, to build synthetic power networks, and find that the power network superfamily is formed with a fewer number of transmission lines. Furthermore, we explain that the power network superfamily can maintain a high level of global synchronization stability at a low construction cost, \textit{i.e.}, the shorter transmission lines statistically measured by network redundancies.

The paper is organized into four sections. 
In Section~\ref{sec:motif}, we discuss the five triadic subgraphs and their significance in typical power networks.
In Section~\ref{sec:superfamily}, we discuss the triad significance profile and the power network superfamily. 
In Section~\ref{sec:stablity}, we discuss the unique network structure of the power network superfamily due to the trade-off between the network synchronization stability and network redundancy. 
The final discussion and concluding remarks are given in Section~\ref{sec:conclusion}.

\section{Subgraph Significance in Power Networks}\label{sec:motif}  
A high-voltage transmission power grid can be modeled as a unidirectional complex network\cite{D?rfler2012} $G(\bm{V},\bm{E},\bm{A})$, where $\bm{V}$ is the set of nodes, $\bm{E}$ the set of edges and $\bm{A}$ the unidirectional adjacency matrix. The node $V_i \in \bm{V}$ represents a power/transformer station in power grid. The edge $E_{ij} \in \bm{E}$ represents the transmission line between nodes $V_i$ and $V_j$. Considering that the electric energy flows one node to the other through the edge, $\bm{A}$ is used to record the unidirectional topology where $A_{ij}=1$ denotes an energy transmission from $V_i$ to $V_j$ in $E_{ij}$, otherwise $A_{ij}=0$ means there is no edge between  $V_i$ and $V_j$ or the direction of power flow in $E_{ij}$ is from $V_j$ to $V_i$. For simplicity, the direction of power transmission in $E_{ij}$ is determined by phases\cite{Filatrella2007} of the two nodes based on the DC model.

\subsection{Triadic Subgraph Significance} \label{sec:z_score_null}
\begin{figure}
   \includegraphics[width=1.0\textwidth]{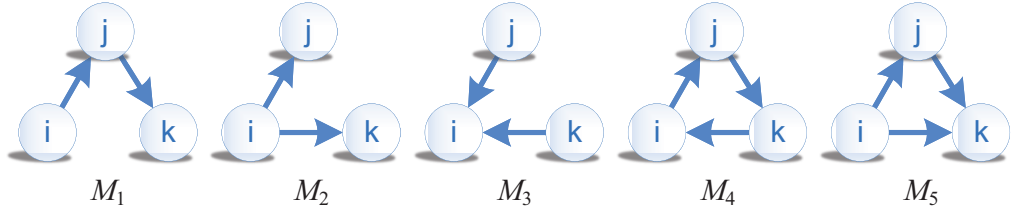}%
   \caption{Five unidirectional triadic subgraphs}
   \label{fig:motif5types}
\end{figure}

Network motifs are recurrent and significant subgraphs and considered building blocks for complex networks\cite{Milo2002}. 
Due to the sparseness of power grids\cite{Schultz2014}, only small local subgraphs could significantly exist in networks. 
Therefore, all the unidirectional triadic subgraphs are used to be found as the candidate motifs in power networks, which are illustrated in Figure~\ref{fig:motif5types}.  
As shown in Figure~\ref{fig:motif5types}, triadic subgraph $M_1$ directs energies from node ${V}_i$ to node ${V}_k$ through node ${V}_j$ along a chain path. 
Triadic subgraphs $M_2$ and $M_3$ deliver/absorb energies in a parallel way. 
Triadic subgraph $M_4$ is known as the feedback loop, and triadic subgraph $M_5$ is the feed-forward loop.

The functional roles of local subgraphs are still unclear in power systems. Although the beneficial function of the subgraphs may not be clear, deviations from null-hypothesis models provide a strong indication that some certain local structures are important to the whole system. To identify which subgraph is significant in power networks, the statistic significance of a triadic subgraph $M$ can be described by the Z score\cite{Milo2004a},
\begin{equation}\label{eq:z_score}
	Z = (N^{grid}-\left\langle  N^{rand}  \right\rangle ) / std(N^{rand}),
\end{equation}
where $N^{grid}$ is the number of $M$ instances in power networks, $\left\langle  N^{rand}  \right\rangle$ and $std(N^{rand})$ stand for the mean and standard deviation of the number of the $M$ instances in null-hypothesis randomized networks. To remove the misleading significance of subgraphs, the null-hypothesis randomized networks are used as the reference networks, which are generated by the following two-step swap scheme,
\begin{enumerate}
\renewcommand{\labelenumi}{Step \theenumi}
\setlength{\itemindent}{2.0em}
\item: Remove two randomly selected unidirectional lines of $A \to a$ and $B \to b$, and create new unidirectional lines $A \to b$ and $B \to a$.
\item: If one of the lines already exists, no swap is carried out, and go back to Step 1.
\end{enumerate}

\begin{figure}
	\centering
	\subfigure[IEEE 57-bus test system]{\includegraphics[width=0.4\textwidth]{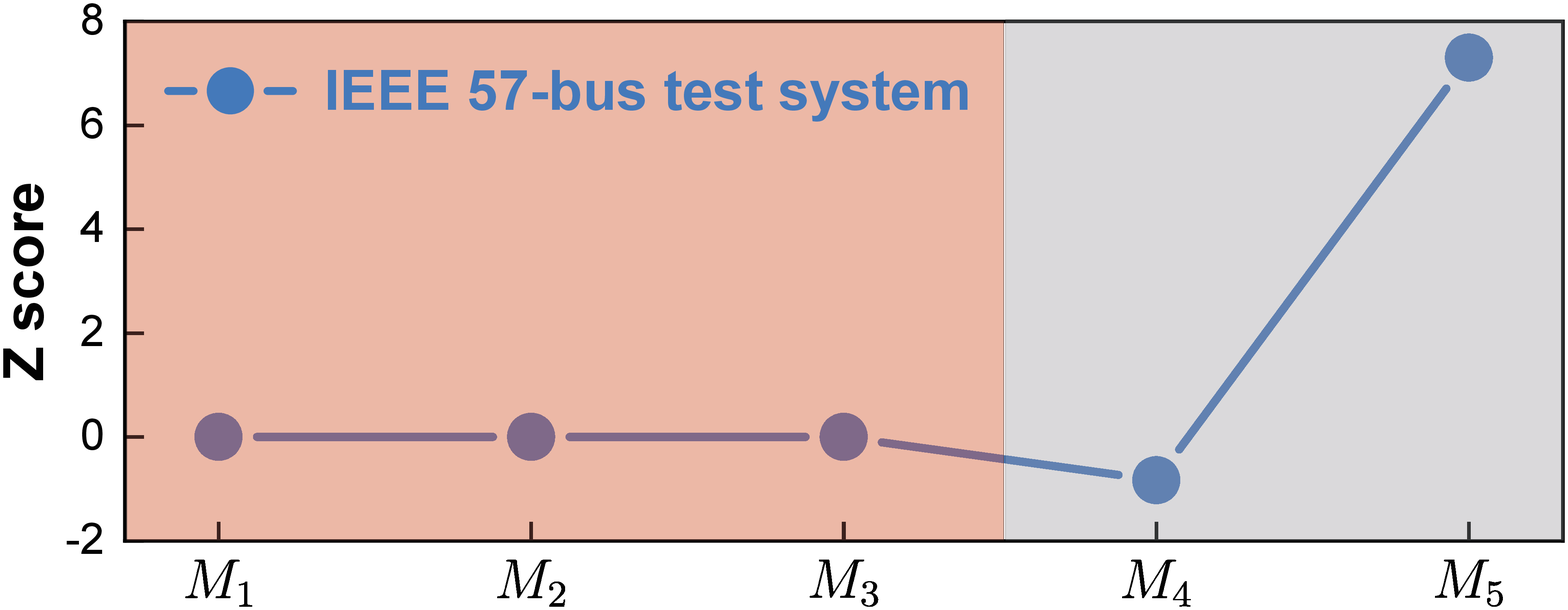}} 
	\subfigure[IEEE 118-bus test system]{\includegraphics[width=0.4\textwidth]{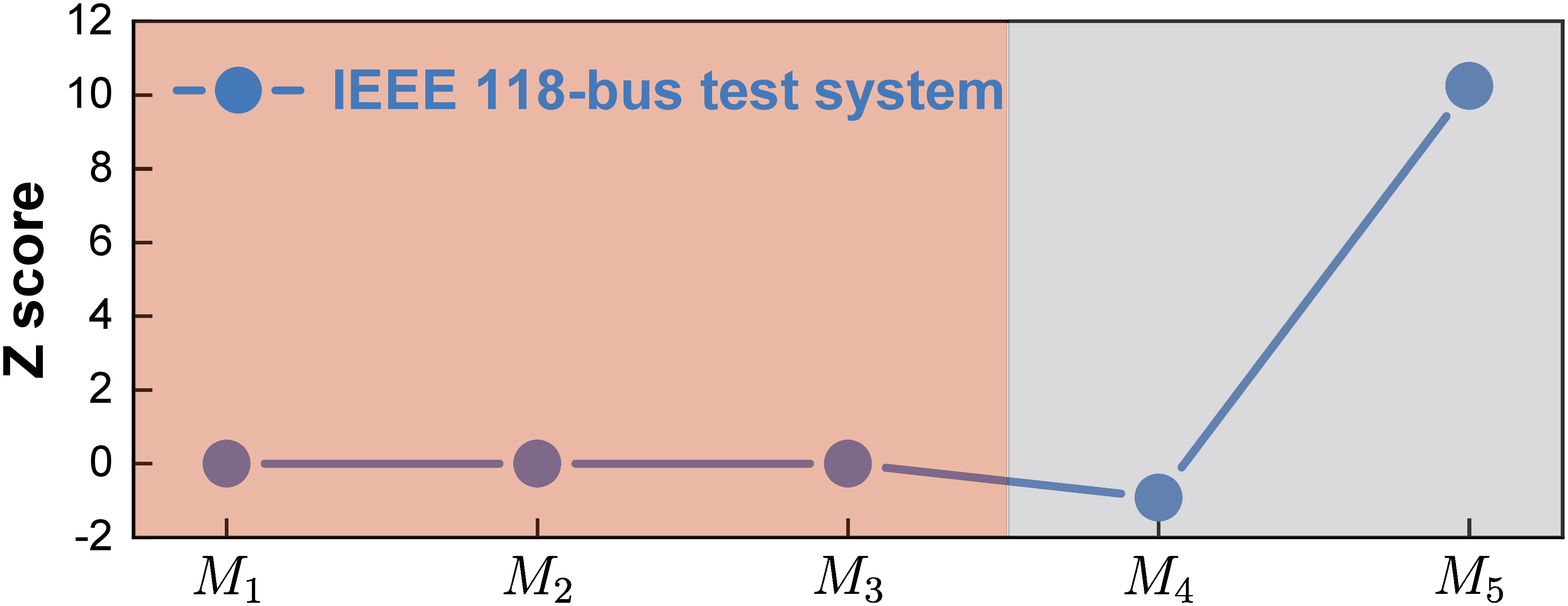}} \\
	\subfigure[IEEE 300-bus test system]{\includegraphics[width=0.4\textwidth]{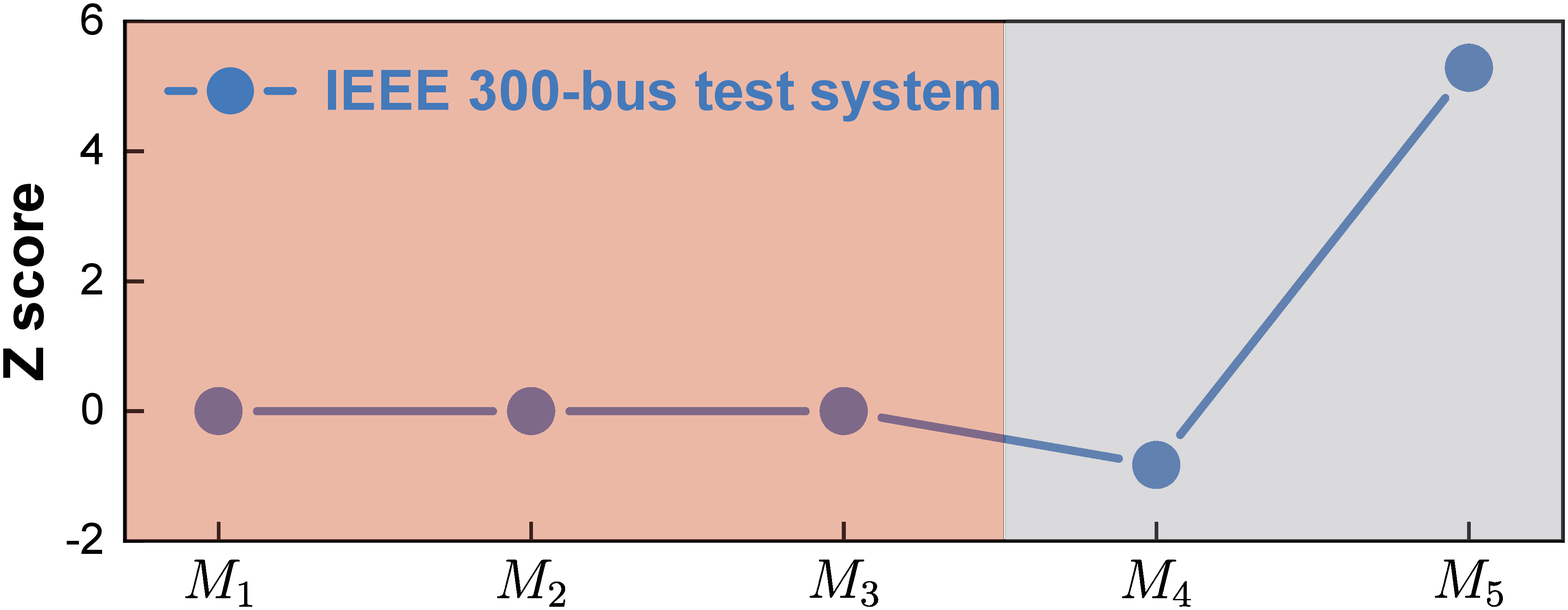}} 
	\subfigure[UK Power Grid]{\includegraphics[width=0.4\textwidth]{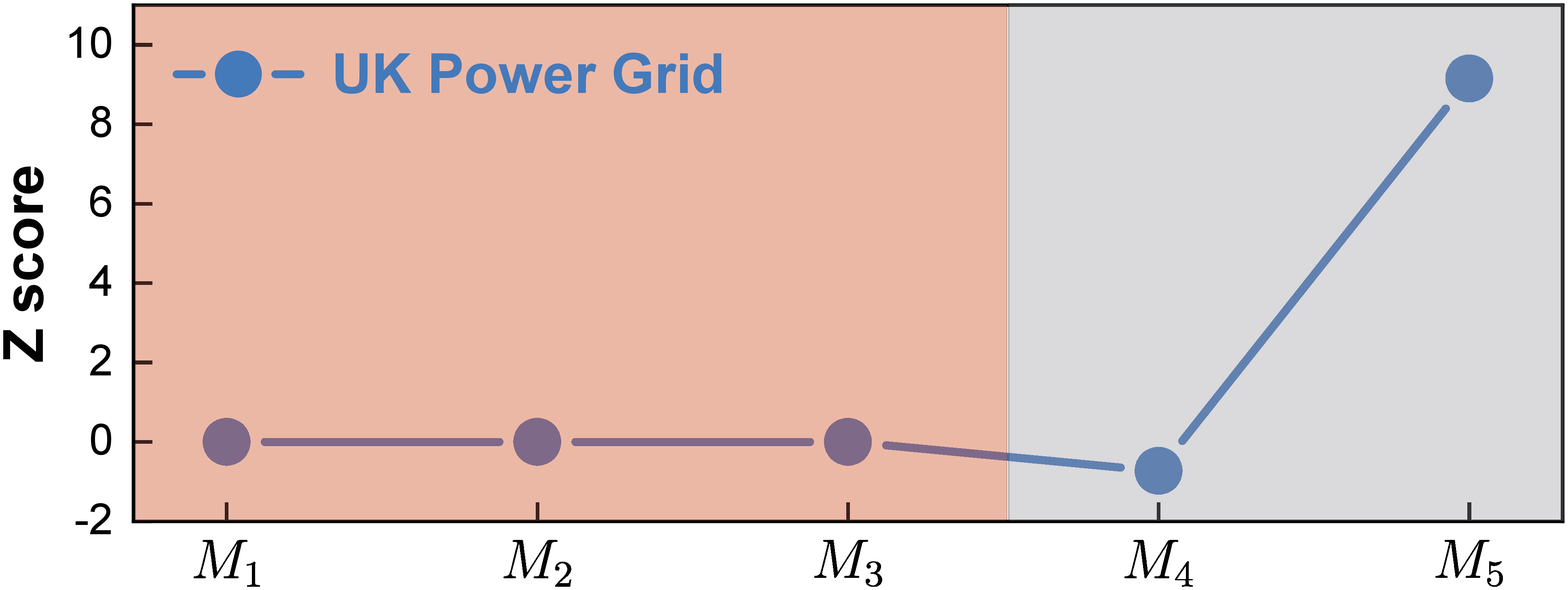}} \\
	\subfigure[North European Grid]{\includegraphics[width=0.4\textwidth]{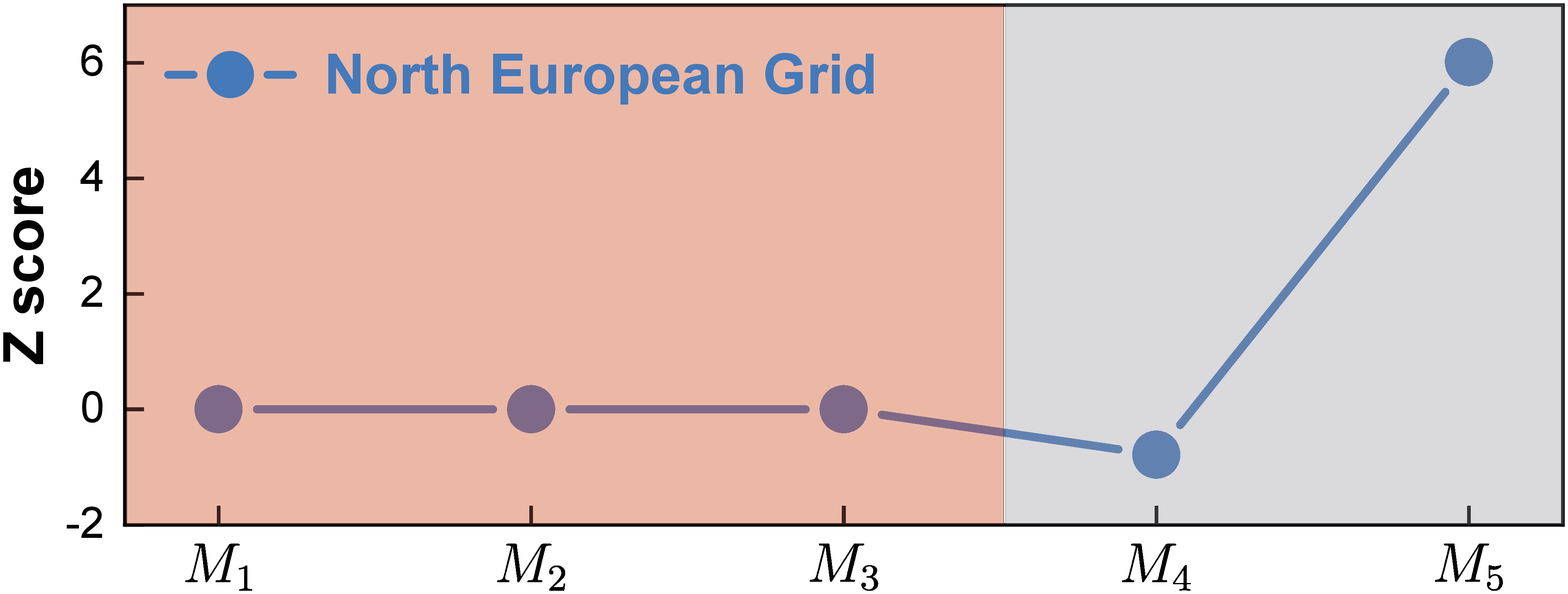}}
	\caption{The Z scores of the five unidirectional triadic subgraphs are calculated for the five typical power networks including IEEE 57-bus test system, IEEE 118-bus test system, IEEE 300-bus test system, the UK Power Grid and the North European Grid. For each power network, 1000 null-hypothesis randomized networks are generated to evaluate Z scores.}%
    \label{fig:super_family_zscore} 
\end{figure}

Since complex networks in the real world are built based on specific design principles and functional constraints, small local subgraphs may have different significance in complex networks\cite{Milo2002,Milo2004a,Benson2016a}. However, the null-hypothesis randomized networks are generated without any design principle or functional constraint. But, these randomized networks maintain the same number of nodes and edges and the same degree sequence as the observed power networks. More specifically, the randomized networks maintain the same lower-order edge-based feature in the observed power networks, such as the number of incoming and outgoing edges for each node. Thus, the significance of local subgraph structures in the observed power networks can be described in comparison with the null-hypothesis randomized networks. We take five typical power networks including IEEE 57-bus test system, IEEE 118-bus test system, IEEE 300-bus test system\cite{Zimmerman2011}, the UK Power Grid\cite{Witthaut2012a} and the North European Grid in Figure~\ref{fig:super_family_zscore}, to evaluate the significance of the five unidirectional triadic subgraphs based on Z scores.
Figure~\ref{fig:super_family_zscore} shows that the five unidirectional triadic subgraphs have the similar relative significance of Z scores for the typical power networks. $Z_1$, $Z_2$ and $Z_3$ are equal to $0$s. $Z_4$ is negative within $(-2, 0)$, and $Z_5$ is positive within $(5, 12)$. 
The observed power networks and null-hypothesis randomized networks have the same number of $M_1$, $M_2$ and $M_3$ instances. 
The $M_4$ triadic closure does not exist and the $M_5$ triadic closure has high significance in power networks. 

\subsection{Zero Z Score in Power Networks}
\begin{figure}
	\centering
	\subfigure[$G_1$ network]{\includegraphics[width=0.45\textwidth]{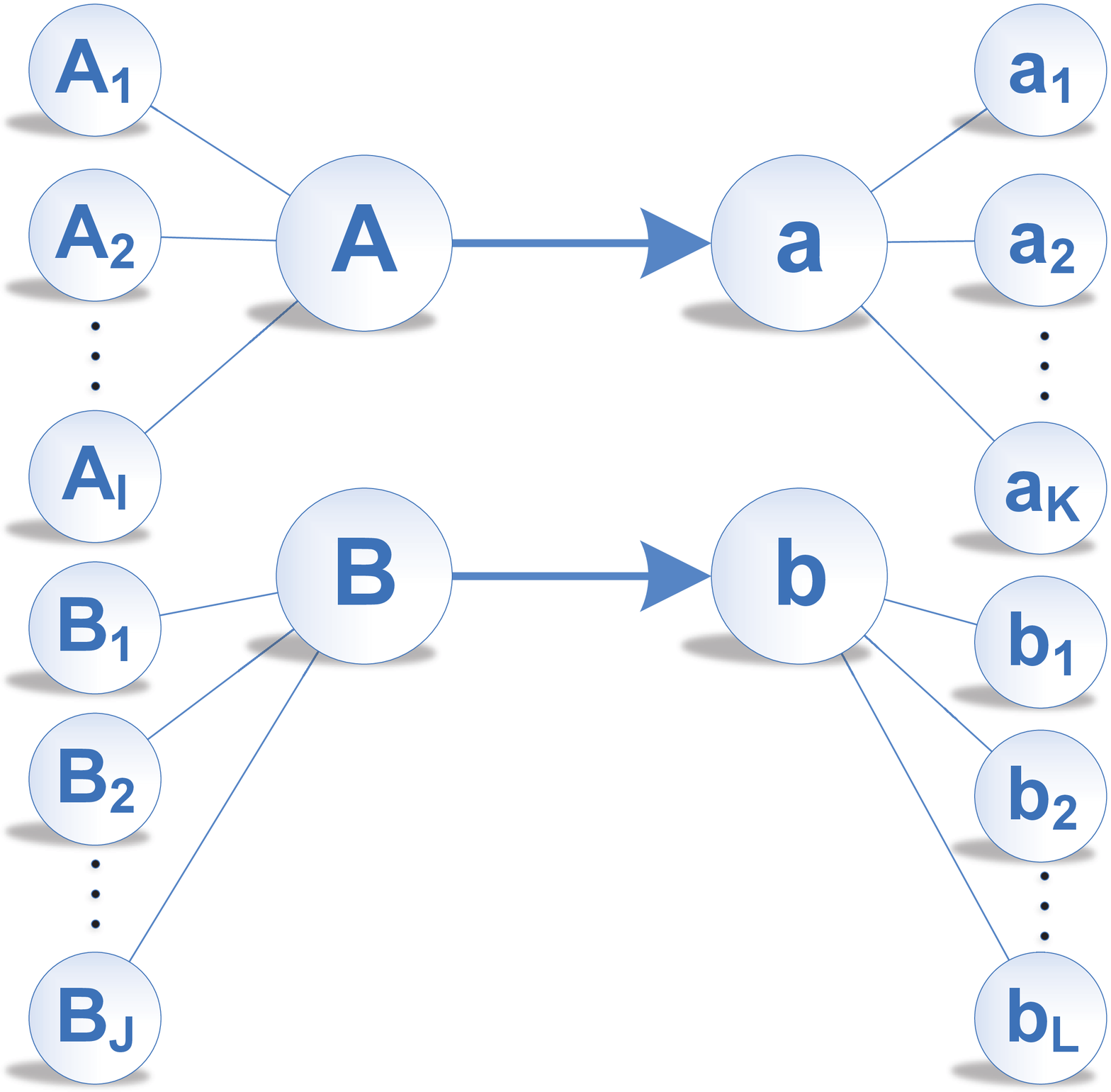}\label{swapa}}
	\subfigure[$G_2$ network]{\includegraphics[width=0.45\textwidth]{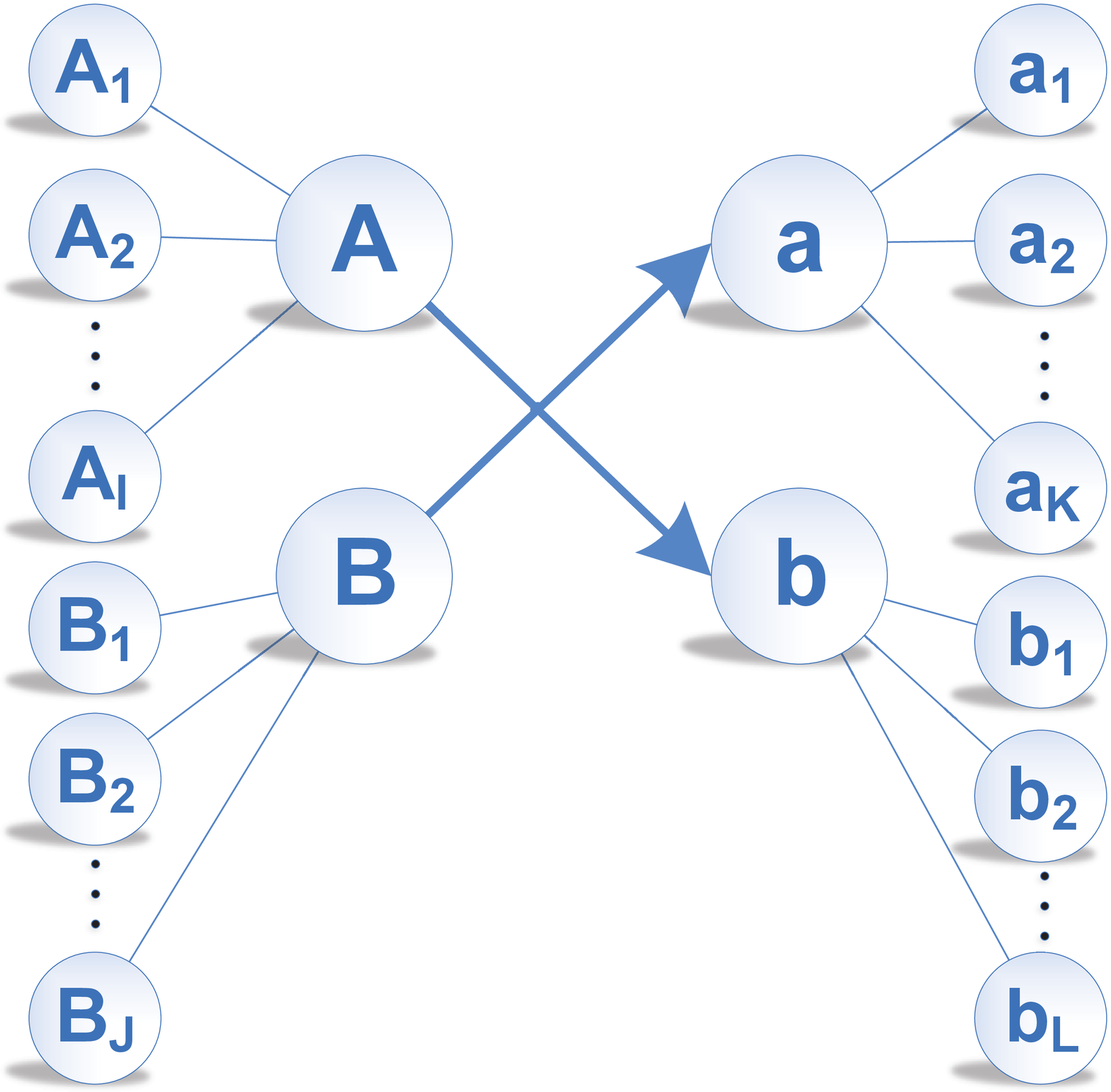}\label{swapb}}
	\caption{The $M_1$, $M_2$ and $M_3$ subgraphs before and after a swap. (a) the two randamly selected unidirectional lines $A \to a$ and $B \to b$ (b) a swap  $A \to a$ and $B \to b$ to $A \to b$ and $B \to a$. A thick arrow represents a unidirectional line `$\to$', and a thin line represents a link with an indefinite direction.}
	\label{fig:swap}
\end{figure}

Z score becomes zero only when $N^{grid}$ in power networks is equal to $\langle N^{rand} \rangle$ in randomized networks. Figure~\ref{fig:swap} shows the swap scheme to explain why subgraphs $M_1$, $M_2$ and $M_3$ have zero Z scores. The randomized networks are still unidirectional due to the swap scheme. As shown in Figure~\ref{swapa}, we randomly choose two lines $A \to a$ and $B \to b$ in a unidirectional network $G_1$. When a swap happens, the two new lines $A \to b$ and $B \to a$ are created and the original two lines $A \to a$ and $B \to b$ are removed, which is shown in Figure~\ref{swapb}. The unidirectional network after the swap is denoted as $G_2$.  
For the $G_1$ and $G_2$ networks, the only difference is the swap links.  
Figure~\ref{fig:swap} shows the number of $M_1$, $M_2$ and $M_3$ instances in the $G_1$ and $G_2$ network is equal according to the swap scheme.
The proof details given below,
\begin{enumerate} 
\item The number of the $M_1$ instances in $G_1$ and $G_2$:
\begin{align}
N(A_{i} \to A \to a|G_1)& =N(A_{i} \to A \to b|G_2),i \in 1,2,\cdots,I, \\ \nonumber
N(B_{j} \to B \to b|G_1)& = N(B_{j} \to B \to a|G_2), j \in 1,2,\cdots,J,  \\ \nonumber
N(A \to a \to a_{k}|G_1)&= N(B \to a \to a_{k}|G_2),  k \in 1,2,\cdots,K,\\ \nonumber
N(B \to b \to b_{l}|G_1)&= N(A \to b \to b_{l}|G_2),  l \in 1,2,\cdots,L.\\ \nonumber
\end{align}
\item The number of the $M_2$ instances in $G_1$ and $G_2$:
\begin{align}
N(A_{i} \gets A \to a|G_1)& = N(A_{i} \gets A \to b|G_2), \\ \nonumber
N(B_{j} \gets B \to b|G_1)& = N(B_{j} \gets B \to a|G_2).\\ \nonumber
\end{align}
\item The number of the $M_3$ instances in $G_1$ and $G_2$:
\begin{align}
N(A \to a \gets a_{k}|G_1) & = N(B \to a \gets a_{k}|G_2), \\ \nonumber
N(B \to b \gets b_{l}|G_1)  &= N(A \to b \gets b_{l}|G_2).\\ \nonumber
\end{align}
\end{enumerate}
where $N(\bullet|\circ)$ represents the number,  $\bullet$ represents triadic subgraph instances, and $\circ$ refers to the $G_1$ network or the $G_2$ network. As a result, the swap scheme can not change the number of $M_1$, $M_2$ and $M_3$ instances. Thus, the Z scores of $M_1$, $M_2$ and $M_3$ subgraphs are exactly zeros in reference to the randomized networks. 

\subsection{Overrepresented Significance of $M_5$ Motif}
The five typical power networks in Figure~\ref{fig:super_family_zscore} all have the underrepresented significance of the $M_4$ triadic closure and the overrepresented significance of the $M_5$ triadic closure. Although the null-hypothesis randomized networks have $M_4$ instances, the power networks have no $M_4$ instances. 

Therefore, to confirm the overrepresented significance of $M_5$ triadic closure in power networks, the alternative null-hypothesis randomized networks without $M_4$ triadic closure are generated by the alternative two-step swap scheme:
\begin{enumerate}
	\renewcommand{\labelenumi}{Step \theenumi}
	\setlength{\itemindent}{2.0em}
	\item: Remove two randomly selected unidirectional lines $A \to a$ and $B \to b$, and create new unidirectional lines $A \to b$ and $B \to a$.
	\item: If one of the lines exists, or any of the $M_4$ instances $\overleftarrow{A \to b \to X}$ or $\overleftarrow{B \to a \to X}$ appears, no swap is carried out and go back to Step 1. Here $X$ represents an arbitary node, and the arrow lines `$\to$' and `$\gets$' represent unidirectional lines in networks.
\end{enumerate}
The alternative swap scheme by automatically discarding $M_4$ instances generates the degree-preserving null-hypothesis randomized networks which obey Kirchhoff's law of the DC model as power networks. 

\begin{figure}
	\centering
	\includegraphics[width=0.8\textwidth]{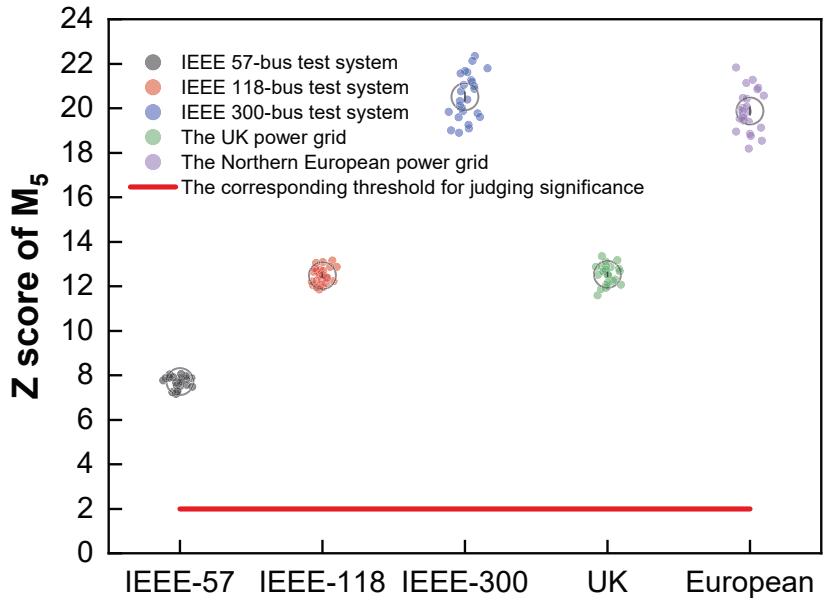}\\
	\caption{The Z scores of $M_5$ triadic closure based on the alternative null-hypothesis model for five typical power networks including IEEE 57-bus test system, IEEE 118-bus test system, IEEE 300-bus test system, the UK Power Grid and the North European Grid. For each power network, the alternative $Z_5$ is calculated 25 times repeatedly, and 1000 alternative null-hypothesis randomized networks with no $M_4$ instances are generated.}
	\label{fig:zscore_no_m4}
\end{figure}

The alternative $Z_1$, $Z_2$ and $Z_3$ are still equal to $0$s in reference to the alternative null-hypothesis randomized networks. Since there exist no $M_4$ instances in the alternative null-hypothesis randomized networks, the alternative $Z_4$ is nonexistent. Therefore, we only evaluate the alternative $Z_5$. Figure~\ref{fig:zscore_no_m4} shows that the alternative $Z_5$ are all very high for the five typical power networks including IEEE 57-bus test system, IEEE 118-bus test system, IEEE 300-bus test system, the UK Power Grid and the North European Grid. Typically, 2 is the threshold of Z score to judge the significance of a subgraph in a network\cite{Milo2004a} as indicated in red line in Figure~\ref{fig:zscore_no_m4}. It is clear that all five typical power networks are far above the threshold. Therefore, the two types of $Z_5$ for the five typical power networks in Figure~\ref{fig:super_family_zscore} and Figure~\ref{fig:zscore_no_m4} both consistently confirm that the $M_5$ triadic closure is significantly overrepresented in power networks in reference to the two null-hypothesis randomized networks. The $M_5$ triadic closure is definitely a motif for power networks. 

Although the $M_4$ triadic closure is significantly underrepresented due to the absence of $M_4$ instances in power networks in reference to the null-hypothesis randomized networks, $Z_4$ can still provide useful information to differentiate the power networks from other complex networks.  
For example, the networks of electronic circuits of digital fractional multipliers\cite{Milo2004a} have the high significance of $M_4$ triadic closure. Thus, with the null-hypothesis randomized networks, we can differentiate power grids from electronic circuits. Although power networks don't have the $M_4$ instances, other complex networks could still have the $M_4$ instances.
The randomized networks are not an ensemble of purely `random' power networks, but networks as a null-hypothesis reference. Since $Z_4$ also contains crucial structural information, the null-hypothesis randomized networks with $M_4$ triadic closure is a more suitable null-hypothesis model to evaluate the significance and reveal the uniqueness of power networks. The next section shows how the significance of $M_4$ and $M_5$ triadic closures makes power networks a network superfamily.

\section{Triadic Local Structures of Power Networks}\label{sec:superfamily}
Z scores of power networks with different sizes are not convenient for comparison. For example, it is hard to say that which is more significant, $Z_5$ in IEEE 118-bus test system or $Z_5$ in IEEE 300-bus test system shown in Figure~\ref{fig:super_family_zscore}. The normalized Z scores of triadic subgraphs, aka the triad significance profile(TSP) \cite{Milo2004a}, are defined as, 
\begin{equation}\label{eq:TSP}
	TSP_{i} = Z_{i} / \left(\sum_{j=1}^{5} Z_{j}^2 \right)^{1/2}.
\end{equation}
The normalization emphasizes the relative significance of these triadic subgraphs among different scale of power grids. 
To keep things simple, we ignore $TSP_1$, $TSP_2$ and $TSP_3$ because all of them are equal to $0$s. 
 
Whether a subgraph plays a crucial part in networks depends on whether it is more likely to develop the subgraph instances in the power network growth. Thus, the significance of a subgraph is essentially characterized by the difference of subgraph instances in power networks from the null-hypothesis randomized networks. 
Generally, the significant difference indicates the subgraph is more important while the minor difference means the subgraph is less important.
In power networks, the triad significance of $M_1$, $M_2$ and $M_3$ subgraphs are always zeros in reference to the randomized networks, which means that $M_1$, $M_2$ and $M_3$ subgraphs may be less important to be ignored.
Although there could still exist small local structures in reality that are functionally important but not statistically significant\cite{Chen2018}, the subgraphs with high significance are more worthy of study.

\begin{figure}
  \centering
  \includegraphics[width=0.9\textwidth]{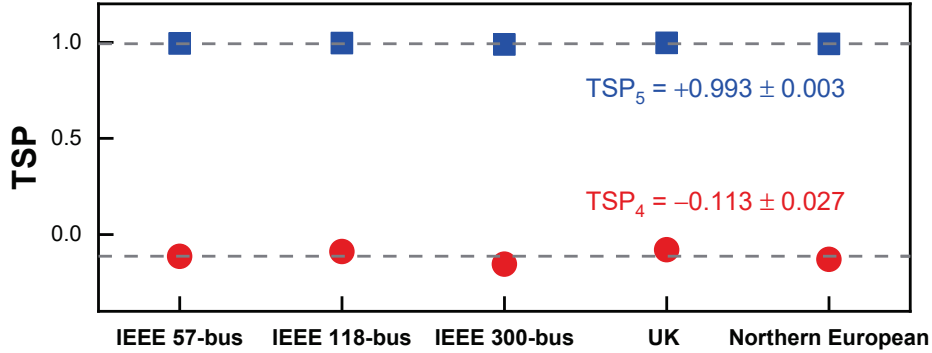}
  \caption{The TSP pair has $TSP_4=-0.113 \pm 0.027$ and $TSP_5=0.993 \pm 0.003$ for the five typical power grids including the IEEE 57-bus, IEEE 118-bus, IEEE 300-bus test system, the UK Power Grid and the North European Grid. The red dots represent $TSP_4$ and the blue squares represent $TSP_5$.}%
  \label{fig:super_family_tsp}
\end{figure}

The $M_4$ and $M_5$ triadic closures have unique significances. The Z score of $M_4$ triadic closure shows the underrepresented significance in power networks while the Z score of $M_5$ triadic closure demonstrates the overrepresented significance. Jointly, Figure~\ref{fig:super_family_tsp} shows the TSP pairs of $M_4$ and $M_5$ triadic closures for the five typical power networks who all have $TSP_4=-0.113$ with a standard deviation of $0.027$ and $TSP_5=0.993$ with a standard deviation of $0.003$. The TSP pair values stay very constant for typical power networks. Power networks form a network superfamily\cite{Menck2014a} at the motif level characterized with the TSP pair.

\subsection{TSP Pair as Identifier of Power Networks}\label{sec:structural_particularity}
The power networks share the TSP pair of $TSP_4=-0.113$ and $TSP_5=0.993$. To further confirm that the TSP pair can server as an identifier of power networks, we compare power networks with the well-known random networks, small-world networks and scale-free networks. The IEEE 118-bus test system with 118 nodes and 179 edges is used as the exemplary power network. For the comparison, we take three typical complex networks, \textit{i.e.}, the typical small-world (SW) networks, the completely random SW networks, and the scale-free (SF) networks. The three typical complex networks have the same nodes and edges as the IEEE 118-bus test system.  

The SW networks\cite{Watts1998} with 118 nodes and 179 edges are generated as the following steps: 
\begin{enumerate}
	\renewcommand{\labelenumi}{Step \theenumi}
	\setlength{\itemindent}{2.0em}
	\item: Create a ring of 118 nodes: Each node in the ring is connected with its two nearest neighbors; Randomly choose different 62 nodes. For each selected node $V_i$, create an edge $E_{i,i+2}$; 
	\item: Create shortcuts by rewiring edges: For each line $E_{uv}$ in the ring, a node $w$ is randomly chosen and a new line $E_{uw}$ is added with rewiring probability $p_{re}$. Once the new line $E_{uw}$ is created, the old line $E_{uv}$ must be removed;
	\item: Check whether the network after rewiring is connected: If the network is not connected, repeat Step 1 and 2 again.
\end{enumerate}
When the rewiring probability $p_{re}=0$, the networks are regular. 
As $p_{re}$ increases, the networks become more random and the small-worldness $\eta \triangleq \left( C/C_r \right) / \left( L/L_r \right) $ changes, where $C$ is clustering coefficient, $C_r$ is the average clustering coefficient of the equivalent random networks with same degree, $L$ is the path length and $L_r$ is the average path length of the equivalent random networks. 
When $p_{re}$ achieves $100\%$, the SW networks become completely random.
In this paper, the typical SW networks are defined as those SW networks with the highest small-worldness $\eta$ (about $p_{re}=18\%$), and the completely random SW networks are those SW networks generated by $p_{re}=100\%$. 

The scale-free (SF) networks\cite{Barabasi1999} with 118 nodes and 179 edges are generated as the following steps:
\begin{enumerate}
	\renewcommand{\labelenumi}{Step \theenumi}
	\setlength{\itemindent}{2.0em}
	\item: Create a random minimum spanning tree with 56 nodes and 55 edges initially;
	\item: Generate a new node to link with two different existing nodes by 2 new lines. The nodes are selected from the existing nodes randomly according to the probability $\Pi\left( w \right) =\frac{deg\left( w\right) }{\sum_{j \in \bm{V}}deg\left( j \right) }$ where $w$ represents the selected node and $deg(\cdot)$ the nodal centrality of degree. Repeat this step 62 times to make the number of existing nodes become 118.
\end{enumerate}

\begin{figure}
	\centering
	\includegraphics[width=0.8\textwidth]{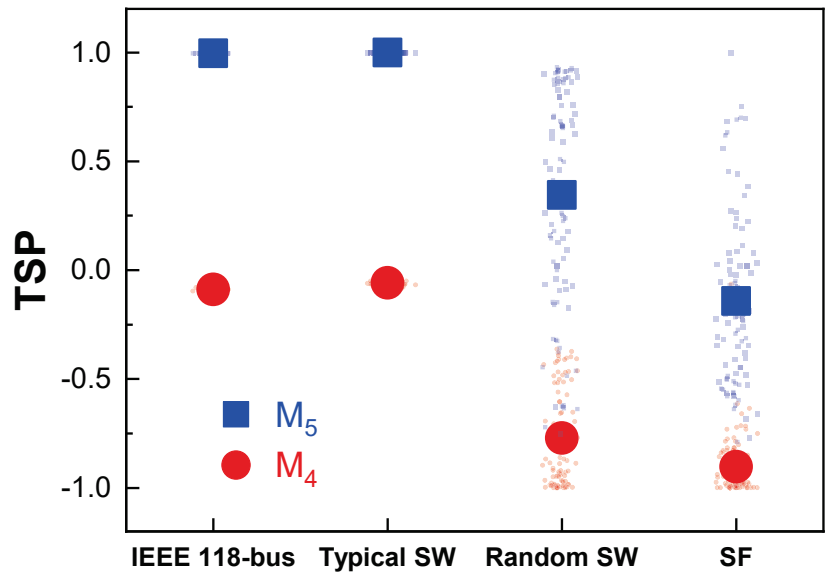}
	\caption{The TSP pairs of IEEE 118-bus test system, the typical SW networks, the completely random SW networks and the SF networks. 100 samples are generated for each type of networks and each sample has a $TSP_4$ value drawn as a small red dot and a $TSP_5$ value drawn as a small blue square. The large red dots represent the average $TSP_4$ and the large blue squares represent the average $TSP_5$ for four types of networks.}%
	\label{fig:sf_sw_118}
\end{figure}

The comparison of the four networks reveals the structural particularity of the power network superfamily. 
Figure~\ref{fig:sf_sw_118} shows the TSP pairs for the IEEE 118-bus test system, the typical SW networks ($p_{re}$ = 18\%), the completely random SW networks ($p_{re}$ = 100\%) and the SF networks. 
The TSP pair for the IEEE 118-bus test system is roughly similar to the pair of the typical SW networks, while much different from the pairs of completely random SW networks and the SF networks. Figure~\ref{fig:sf_sw_118} implies that power networks have small-worldness to a certain extent. The design principles of power systems may favor the small-worldness of power grids, \textit{i.e.}, the need for short path length between two nodes far away geographically, N-1 security criterion to promote loop structures, and diminishing dead tree structures\cite{Menck2014a}. So the power networks cannot be completely random. Furthermore, since the hub nodes and numerous leaf structures can be weak points in power systems, the power networks cannot be scale-free either. 

\subsection{Synthetic Power Networks in Power Network Superfamily}\label{sec:RG}
For the typical SW networks, their TSP pair has $TSP_4=-0.058$ and $TSP_5=0.998$ in Figure~\ref{fig:sf_sw_118}, but the TSP pair of power network has $TSP_4=-0.113 \pm 0.027$ and $TSP_5=0.993 \pm 0.003$. Therefore, although the typical SW networks have similar structures with the power networks to some extent, power networks still have their unique TSP pair at the triadic subgraph level. With the random growth (RG) model\cite{Schultz2014} to generate the synthetic power networks, we can see how a network grows to a power network at the triadic subgraph level.
The growth mechanisms in the RG model are found in real-world power grids. 
Therefore, the unique TSP pair can tell whether the RG model reproduces a similar structure of the real-world power network. Regarding the growth of power networks, the RG model has been proposed\cite{Schultz2014} to generate synthetic power networks through two phases, an initialization phase of creating a minimum spanning tree network and a growth phase of adding nodes and redundant lines. There are five key parameters for the RG model, which are denoted as $\left( n_0, p, q, r, s \right) $, where $n_0$ is the number of nodes to initialize a minimum spanning tree network, $p$ is the probability of adding a redundant line between an existing node and a new node at each growth step, $q$ is the probability of adding a redundant line between two existing nodes, $r$ is the exponent for the cost-vs-redundancy trade-off, and $s$ is the probability splitting a line in each growth.

\begin{figure}
	\centering
	\includegraphics[width=0.9\textwidth]{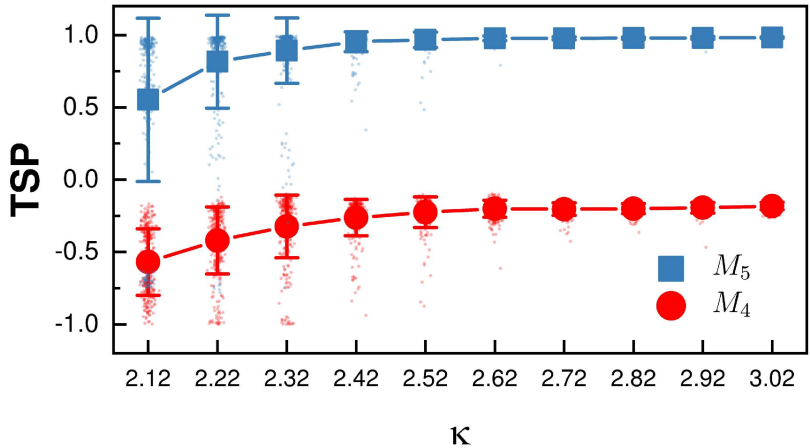}
	\caption{The TSP pairs of $M_4$ and $M_5$ for the synthetic power networks of different average degrees, $\kappa$. 300 sample synthetic power networks are generated for each $\kappa$. 
	}%
    \label{fig:rg_pq}
\end{figure}

We generate synthetic power networks with 100 nodes, whose average degrees range from sparse to dense with a broad range of average degree $\kappa$ with the following parameters for the RG model. The parameters $p$ and $q$ control the network average degree $\kappa$. With $p+q \in [0.06, 0.11, 0.16, 0.21, 0.26, 0.31, 0.36, 0.41, 0.46, 0.51]$, $\kappa$ can range from $2.12$ to $3.02$ since $\kappa =2 + 2p + 2q$. The remaining three parameters are $n_0 = 10$, $r=1$, and $s=0$ which are determined according to the Western US power grid\cite{Schultz2014}. Figure~\ref{fig:rg_pq} shows that as $\kappa$ increases, the synthetic power networks grow larger and denser and approach the realistic power grids gradually in terms of the TSP pair. When $\kappa$ is larger than $2.62$, the change of the TSP pair is not apparent anymore. Empirically, $\kappa$ of the realistic power grids\cite{Pagani2013} is around $\kappa = 2.8$, which is close to $2.62$. Larger $\kappa$ indicates more transmission lines in power grids with fixed nodes. Since the power network superfamily is characterized by its unique TSP pair, Figure~\ref{fig:rg_pq} tells that a network grows and approaches the structure belonging to the power network superfamily with a fewer number of transmission lines. 

\section{Network Synchronization Stability and Significance of Triadic Closure}\label{sec:stablity}
The TSP pair is a higher-order structural feature that differentiates power networks from other complex networks, as discussed in Section~\ref{sec:superfamily}. 
The $M_5$ subgraph is the only existing triadic closure in power networks due to the absence of $M_4$ triadic closure in power grids.  
The significance of $M_5$ triadic closure in power networks is impressively high, which means $M_5$ is a motif favored by the power network growth. However, the functional importance of the $M_5$ triadic closure is still unclear in the whole power network. Intuitively, $M_5$ triadic closure is redundant because it can be reduced to subgraphs $M_1$, $M_2$, and $M_3$ by removing any edge without loss of network connectivity. But, on the other hand, the performance of power networks is strongly correlated with the $M_5$ triadic closure. We systematically generate the complex networks with the different significance of $M_5$ triadic closure and establish the relationship between the motif-based structures and network synchronization stability. 

\subsection{Triadic Closure Significance and Power Network Stability} 
Power systems have multiple synchronous states for the transient dynamics of power systemss\cite{Dorfler2013,D?rfler2012} according to the swing equation in Eq.~\ref{eq:swing_kuramoto}, 
\begin{equation}\label{eq:swing_kuramoto}
\ddot{\theta_{i}}=-\alpha_{i}\dot{\theta_{i}}+p_{m,i}- \sum_{j=1}^{N}K_{ij} \sin \left( (\theta_{i}^{*}-\theta_{j}^{*})+(\theta_{i} - \theta_{j}) \right),
\end{equation}
where N is the number of nodes in power grid, $\theta_{i}$ is the phase difference from the operating synchronous steady state of node $V_i$ in a reference frame co-rotating at the rated frequency.
The operating synchronous steady state of the system is denoted as $(\bm{\theta^*,\bm{\dot{\theta}^*}=\bm{0}})$, where $\bm{\theta^*}$ represents $[\theta_1^{*}, \cdots, \theta_N^{*}]$ and $\bm{\dot{\theta}^*}=\bm{0}$ means $[\dot{\theta}^*_1=0,\cdots,\dot{\theta}^*_N=0]$. 
$\alpha_{i}$ is the damping coefficient of node $V_i$. 
$p_{m,i}$ represents the power injection of node $V_i$, and $K_{ij}$ is the capacity of transmission line $E_{ij}$. 

\begin{figure}
  \centering
  \subfigure[]{\includegraphics[width=0.45\textwidth]{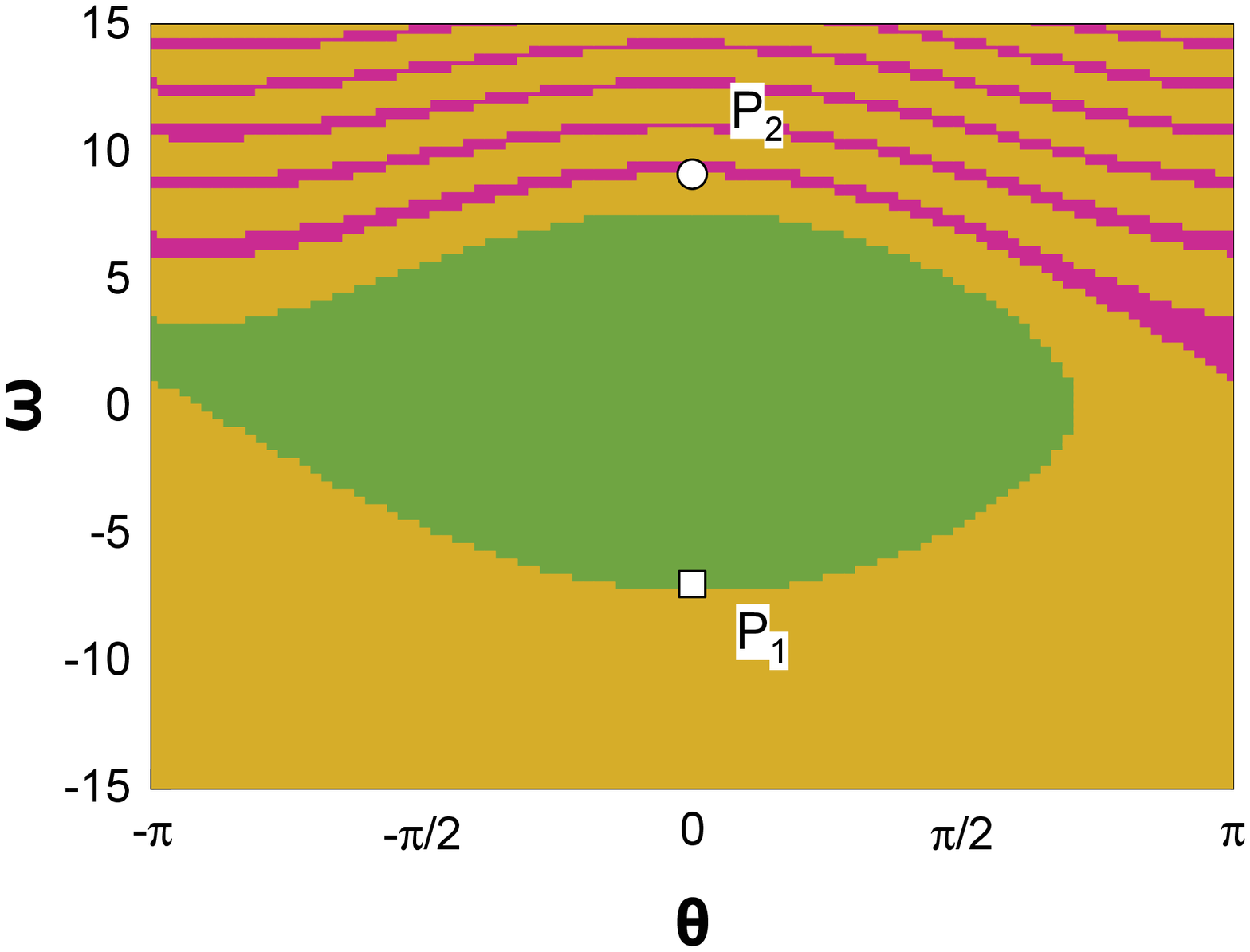} \label{fig:basin_areas1}}
  \subfigure[]{\includegraphics[width=0.45\textwidth]{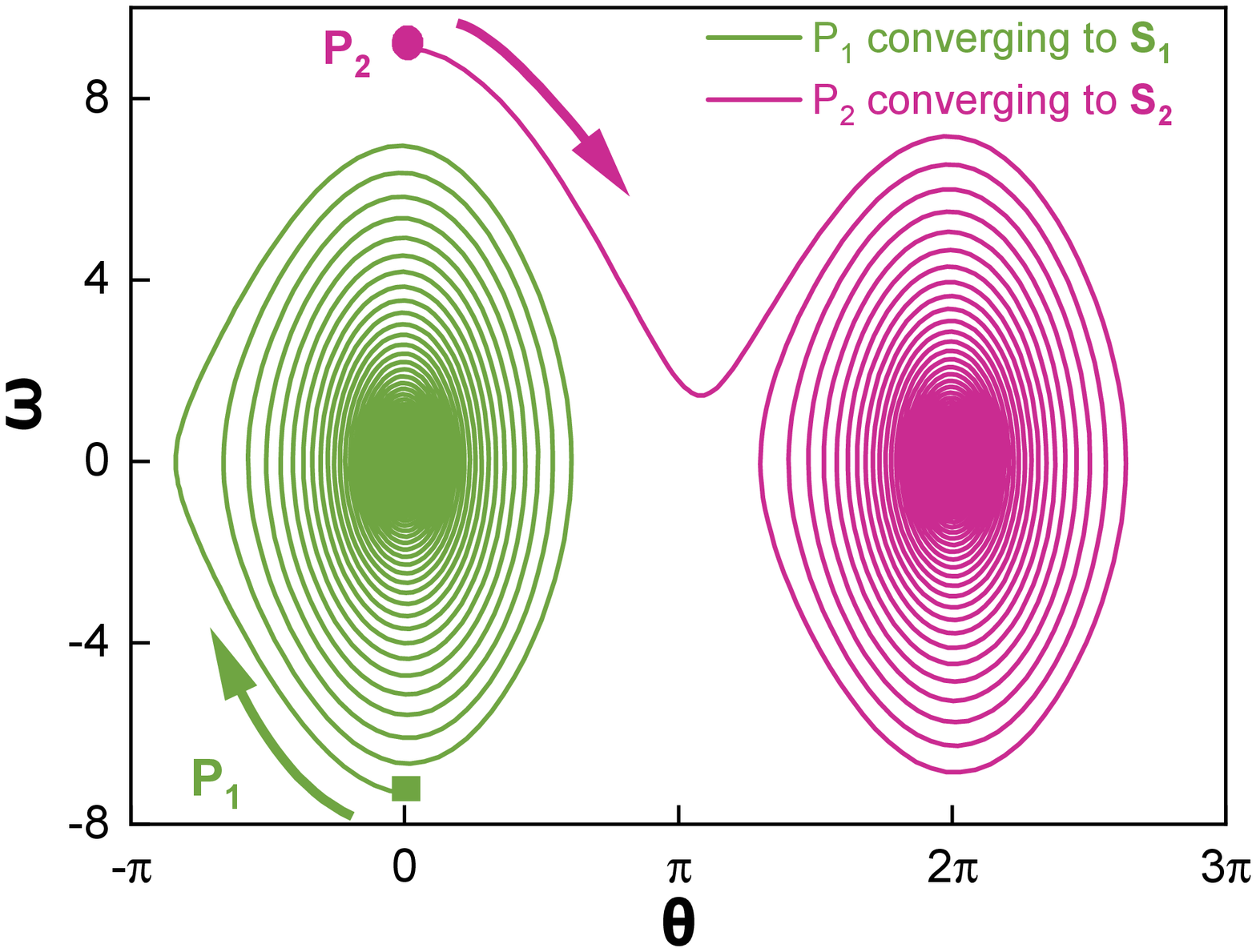} \label{fig:basin_areas2}} 
  \caption{The basin stability of the load node in a two-node power system. 
  (a) The phase diagram has three regions. The green region is the volume of the $S_{1}$ attractor, the red region the volume of the $S_2$ attractor and the orange region the volume of the initial conditions for the unstable states. (b) The green curve is the trajectory of the $P_1$ initial state converging to the $S_{1}$ attractor, and the red curve is the trajectory of the $P_2$ initial state converging to the $S_{2}$ attractor.}
  \label{fig:basin_areas}
\end{figure}

Basin stability\cite{Menck2013} is a measure of global synchronization stability related to the volume of attraction in state space.
Figure~\ref{fig:basin_areas} shows the basin stability of the load node in a uniform two-node power system. 
The generator node $V_G$ has a power input $p_{G}=+1.0$, and the load node $V_L$ has a power output $p_{L}=-1.0$. The coupling strength is $K=8$, and the damping coefficient is $\alpha=0.1$ for two nodes. 
The initial states of the node phase and frequency are drawn uniformly from the state space $\mathcal{Q}= [-\pi,\pi]\times[-15,+15]$.
Figure~\ref{fig:basin_areas1} shows the basin stability of the load node. There are three different regions. The green region is the basin of attraction for the operating synchronous state, denoted as $S_1$ attractor. The red region is the basin of attraction for the alternative synchronous states, denoted as $S_2$ attractor. And the orange region is the volume of initial conditions for the unstable states.  In Figure~\ref{fig:basin_areas2}, the trajectories with two different initial states of $P_1$ and $P_2$ evolve to the two attractors of $S_1$ and $S_2$ respectively. It should be emphasized that power grids may have multiple synchronous states when the scale of the network is large, but all the trajectories can always be classified into two categories, $S_1$ and $S_2$ attractors shown in Figure~\ref{fig:basin_areas}.
The volume of both $S_1$ and $S_2$ attractors contributes to the node synchronization stability in power grids. 

The network synchronization stability is measured with the average of the basin stability\cite{Menck2013} for all nodes in network as $\left\langle BS\right\rangle  =\frac{1}{N} \sum_{i=1}^{N} BS_i$, where $BS_i= \mathcal{B}_i / \mathcal{Q}$ represents the basin stability of node $V_i$. $\mathcal{Q}$ is the volume (number) of the initial conditions drawn from the common state space, and $\mathcal{B}_i$ is the volume (number) of the initial conditions for node $V_i$ that can converge to the synchronous attractors. 

\begin{figure}
	\centering
	\includegraphics[width=0.9\textwidth]{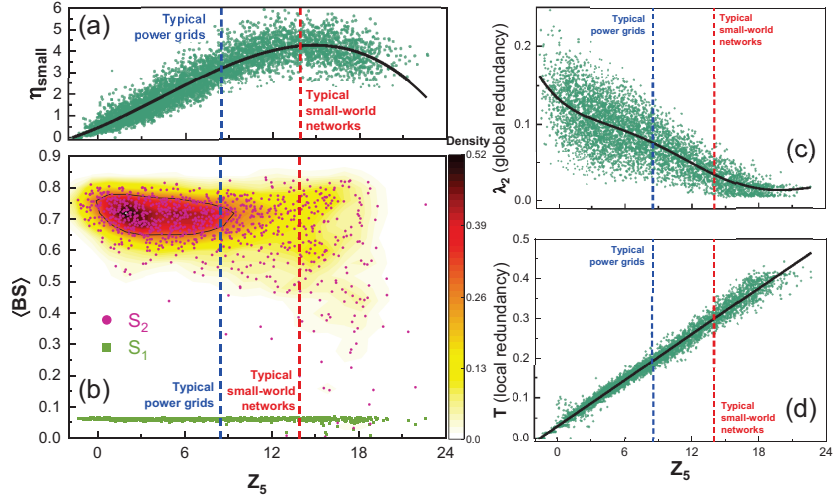}
	\caption{The impact of significance of $M_5$ triadic closure on network synchronization stability and structural properties in SW networks. (a) Small-worldness, (b) network basin stability $\left\langle BS \right\rangle$, (c) algebraic connectivity $\lambda_2$ (global redundancy) and (d) transitivity $\bm{T}$ (local redundancy) {\it{v.s.}} the significance of $M_5$ triadic closure, $Z_5$.
	The blue dashed line represents the typical power grids with the unique TSP pair or $Z_5 = 8.5$.
	The red dashed line represents the typical small-world networks with $Z_5 = 13.9$.
	In panels (a), (c) and (d), each dark green dot represents a sample network generated by the SW model. The rewiring probability $p_{re}$ ranges from $[0,2\%,4\%,\cdots,98\%,100\%]$, and for each $p_{re}$, 100 sample networks are randomly generated. The black curves are obtained by the 4th degree polynomial fitting with R-Square = 0.86 in (a), the 4th degree polynomial fitting with R-Square = 0.65 in (c) and the linear fitting with R-Square = 0.98 in (d). 
	Each light green square dot in panel (b) represents $S_1$ basin stability, and each red dot represents $S_2$ basin stability for a network generated by the SW model. The smoothed color areas represent the underlying probability density of generating the networks with a certain $\left\langle BS \right\rangle$ and $Z_5$. As the color turns darker, the probability density becomes higher. The red area circled by the black contour line has a probability density larger than 0.26, the half of the largest probability density.
	}
    \label{fig:basin_sw} 
\end{figure}

Figure~\ref{fig:basin_sw}(a) shows the network small-worldness $\eta_{small}$\cite{Watts1998} for different $Z_5$.  

The SW networks with different small-worldness show different significance of $M_5$ triadic closure through adjusting the rewiring probability $p_{re}$.The SW model can partially reflect the evolutionary mechanism of power grids due to the similarity of TSP pairs between the IEEE 118-bus test system and the typical SW networks in Figure~\ref{fig:sf_sw_118}. The SW model is used to generate complex networks from regular ($p_{re}=0$) to random ($p_{re}=100\%$), which may contain the typical power networks with a certain rewiring probability ($0<p_{re}<100\%$). 
The power network in the real world often has a limited number of edges to connect with nodes due to the constraint of construction cost. For simplicity, all the SW networks with 60 nodes and 81 edges ($\kappa=2.7$) have the same average degree as the Northern European power grid and the UK power grid. For all the SW networks, their dynamics in Eq.~\ref{eq:swing_kuramoto} have the same parameters. 12 nodes are randomly selected as generators with $p_m=+4$ and the remaining 48 nodes are consumers with $p_m=-1$. The damping coefficient is $\alpha=0.2$ for each node and the coupling strength is $K=8$ for each edge.

The typical small-world networks are the networks with the highest small-worldness $\eta_{small}$ and the typical power grids  are the networks with the unique TSP pair for power network superfamily. To better understand the organization of power networks, the local and global redundancy $\it{vs.}$ $Z_5$ are plotted in Figures~\ref{fig:basin_sw}(c) and (d). 
The global redundancy in Figure~\ref{fig:basin_sw}(c) is measured with algebraic connectivity and the local redundancy in Figure~\ref{fig:basin_sw}(d) is measured with transitivity.
The global redundancy is associated with the network synchronization stability, and the local one is associated with the robustness against cascading failures\cite{Plietzsch2016}. The typical power networks clearly have lower $Z_5$ than the typical small-world networks, but longer average shortest paths. Therefore, intuitively, power grids not only have enough local redundancies, {\it{i.e.}} $M_5$ instances against cascading failures but also have enough global redundancies, {\it{i.e.}} long-range transmission lines to maintain frequency stability\cite{Plietzsch2016}. 

Figure~\ref{fig:basin_sw}(b) shows the $S_{1}$ and $S_{2}$ basin stabilities of 1000 sample SW networks randomly drawn from the networks in Figure~\ref{fig:basin_sw}(a). Since $S_{1}$ basin stability remains constant with different $Z_5$, the significance of $M_5$ triadic closure has little impact on the $S_{1}$ basin stability. Theoretically, the non-zero maximum Lyapunov exponents based on Eq.~\ref{eq:swing_kuramoto}\cite{Motter2013,Menck2014a} can be used to measure the $S_{1}$ basin stability for networks because the $S_{1}$ basin of attraction is close to the operating synchronous state and its volume is small. For the SW networks in Figure~\ref{fig:basin_sw}(b), their non-zero maximum Lyapunov exponents are all $-\frac{\alpha}{2}=-0.1$, which can be used to explain the invariance of $S_{1}$ basin stability. But, for most cases, the $S_{2}$ basin stability determines the network synchronization stability because the $S_{1}$ basin stability is too small, and the $S_{2}$ basin stability dominates. The smoothed color areas of $S_{2}$ basin stability represent the probability density distribution of $\left\langle BS \right\rangle$ and $Z_5$ for the SW networks. The color density is calculated by Bivariate Kernel Density Estimator\cite{Botev2010}. Figure~\ref{fig:basin_sw}(b) shows that the probability density distribution of $S_2$ basin stability become more scattered with the increasing $Z_5$. In other words, the probability of finding networks with large basin stability decreases as $Z_5$ rises. In Figure~\ref{fig:basin_sw} (b), the largest density of network occurrence probability is 0.52 estimated by all the sampling networks. The solid black line indicates the boundary of the area of high probability density larger than 0.26, half of the largest density. The typical power grid is on the blue dashed line $Z_5=8.5$, which intersects with the right edge of the area of high probability density. 
To be more specific, on the right side of the blue dashed line representing typical power network topology, the distribution of $S_2$ basin stability concentrates around $0.7$. But on the left side of the blue dashed line, the distribution of $S_2$ basin stability is scattered significantly as $Z_5$ increases. The $Z_5$ of the typical power network is the dividing point. 
The networks on the left of the blue dashed line will have fewer $M_5$ triadic closures and lower local redundancy, and more long-range transmission lines, and higher global redundancy than typical power grids as shown in Figures~\ref{fig:basin_sw}(c) and (d). When the number of nodes and edges are fixed, more long-range transmission lines and fewer local triadic closures indicate that longer transmission lines are needed to connect with power nodes, which means the power grid has a higher cost. However, building more long-range transmission lines cannot promote the network basin stability at all for the typical power networks from Figure~\ref{fig:basin_sw}(b). 
On the other hand,  the power network on the right of the blue dashed line will have fewer long-range transmission lines and more $M_5$ triadic closures. Thus, although the power grid will have less construction cost given fewer long-range transmission lines, it will risk diminishing network basin stability. According to the relationship of $Z_5$ and networt redundancy shown in Figures~\ref{fig:basin_sw}(c) and (d), the contruction cost can be measured with network redundancy.  
Therefore, the network structures characterized by the unique TSP pair are the trade-off that allows power networks to accomplish the high network basin stability at a low construction cost. The uniqueness of power networks is the balance of network stability and construction cost. 

\section{Discussion and Concluding Remarks}\label{sec:conclusion}
Triadic subgraph analysis reveals the structural features in power networks based on higher-order connectivity patterns.  Five unidirectional triadic subgraphs are identified in power grids.  The triad significance profiles (TSP)  of the five subgraphs are estimated compared to the randomized networks. As a result, power grids demonstrate the unique TSP and form a network superfamily.  We compare the power networks to small-world, scale-free and random networks to verify the uniqueness of TSP for the power network superfamily.
Furthermore, we use the random growth model to generate synthetic power networks to understand the power grids. When the synthetic power networks get denser and their average degree is beyond 2.62, their structures will approach the real-world power networks and fall into the power network superfamily in terms of TSP. Note that the real-world power grids are complex networks whose average degree is 2.8 statistically, close to the threshold value of 2.62 of the synthetic power networks. In other words, the real-world power grid in the superfamily has the least dense network structure as well as the fewest transmission lines. 

From the triadic subgraph perspective, power grids have optimized network structures to balance synchronization stability and network redundancy. Statistically, the network redundancy is related to the construction cost. Therefore, the local subgraph structures of complex networks influence the evolution of power grids. In this paper, the structural features based on higher-order connectivity patterns are revealed and explained by the performance of global synchronization for power systems. The unique network structure in the power network superfamily allows power networks to maintain high synchronization stability at a low construction cost.

For the most artificial complex systems, the trade-off between performance and cost is always a vital issue. Therefore, a better understanding of how the subgraph structures influence the behaviors of complex networks can help design resilient and stable power systems. In the future,  discovering the higher-order organization in power complex networks at the level of small network subgraphs demands more work. It can allow us to understand the higher-order organization's functional role in the power networks. It will be instrumental in optimizing complex power networks and making them more robust and stable.

\section{Acknowledgments}
Xin Chen acknowledges the funding support from the National Natural Science Foundation of China under grant No. 21773182 and the support of HPC Platform, Xi’an Jiaotong University.

Hao Liu and Xin Chen contributed equally to this work.

\section{Data Availability Statement}
The data about the IEEE 57-bus test system, IEEE 118-bus test system, and IEEE 300-bus test system of this study are openly available in MATPOWER at https://matpower.org/, reference number \cite{Zimmerman2011}.
The data about the UK Power Grid and the North European Grid are available in references \cite{Witthaut2012a}, and \cite{Menck2014a} respectively of this study.


\end{document}